\documentclass[12pt]{iopart} 

\usepackage[utf8]{inputenc}
\usepackage{graphicx}
\usepackage{epstopdf}

\begin{document}

\title[Exact predictions from Edwards ensemble vs. realistic simulations]{Exact predictions from Edwards ensemble vs. realistic simulations of tapped narrow two-dimensional granular columns}

\author{Ramiro M. Irastorza$^{1}$, C. Manuel Carlevaro$^{1,2}$, Luis A. Pugnaloni$^{3}$}

\address{$^1$ Instituto de F\'isica de L\'iquidos y Sistemas Biol\'ogicos (CONICET La Plata, UNLP), Calle 59 Nro 789, 1900 La Plata, Argentina.\\
$^2$ Universidad Tecnol\'ogica Nacional - FRBA, UDB F\'{\i}sica, Mozart 2300, C1407IVT Buenos Aires, Argentina.\\ $^3$ Dpto. de Ingenier\'ia Mec\'anica, Facultad Regional La Plata, Universidad Tecnol\'ogica Nacional, Av. 60 Esq. 124, 1900 La Plata, Argentina.}
\ead{luis.pugnaloni@frlp.utn.edu.ar (L A Pugnaloni)}

\maketitle

\begin{abstract}
We simulate via a Discrete Element Method the tapping of a narrow column of disk under gravity. For frictionless disks, this system has a simple analytical expression for the density of states in the Edwards volume ensemble. We compare the predictions of the ensemble at constant compactivity against the results for the steady states obtained in the simulations. We show that the steady states cannot be properly described since the microstates sampled are not in correspondence with the predicted distributions, suggesting that the postulates of flat measure and ergodicity are, either or both, invalid for this simple realization of a static granular system. However, we show that certain qualitative features of the volume fluctuations difficult to predict from simple arguments are captured by the theory. 
\end{abstract}

\section{Introduction}

The description of the states that a pack of macroscopic objects can take has become an important subject of debate in Physics since basic equilibrium statistical mechanics was proposed as a suitable framework to deal with this problem \cite{edwards}. It is expected that macrostates consisting of a large collection of configurations (or microstates) compatible with given macroscopic variables that describe the macroscopic state can be readily defined and that the preferred macrostate for the system under given constraints would be predictable as the mean of a suitable distribution of such microstates.

The aforementioned collection of microstates have to be generated with a well defined protocol of repeated perturbations applied to the packing. Such perturbations are necessary since macroscopic objects (such as the grains of a granular sample) interact through non-conservative forces (inelastic collisions and friction) and so dissipate all kinetic energy at the particle scale. Hence, to move from one configuration to another, an external input of energy is mandatory. Whether a given protocol of perturbation leads to a collection of microstates that can be defined as an equilibrium macrostate in the sense described by Thermodynamics is still a matter of debate. Nonetheless, there exist some consensus that in some cases this may be the case. For instance, annealed tapping protocols are known to give reproducible volumes (and its fluctuations) for a given tap amplitude and tap duration \cite{nowak,ribiere}. However, the question remains as to what are the macroscopic variables that fully describe the 
macroscopic state \cite{pugnaloni2010,pugnaloni2011}. 

Volume fluctuations present a controversial feature. Plotted as a function of volume, fluctuations display a maximum at an intermediate volume between the maximum and minimum volume reached by the macrostates in some studies \cite{pugnaloni2010,pugnaloni2011}. This is in contrast with results from experiments using fluidized beds \cite{schroter} (they show a minimum in fluctuations) and experiments with non-jammed packings \cite{puckett} (they present a monotonic decrease of fluctuations with increasing packing fraction).

In this work we exploit a model recently introduced by Bowles and Ashwin \cite{ashwin2009,bowles} that allows for a full analytic calculation of the density of states of a frictionless granular packing. The model has the advantage of corresponding to a realistic representation of an experimentally realizable system and then allows for a validation of the ability of the theory of ensembles to describe steady states attained by real granular samples.

Under Edwards proposal, in the volume ensemble, all possible configurations compatible with a given macroscopic volume are equally probable. However, the configurations sampled in the lab by using external perturbations cannot be set at a prescribed volume. Therefore, it is generally assumed that the collection of states generated by a repeated perturbation, after any transient has fade, should be compatible with a ``canonical ensemble'' of constant compactivity (the analogue of temperature). We generate such collections of microstates by tapping at different intensities using realistic simulations of the Bowles--
Ashwin model. Then, we compare the results with the prediction of the ``canonical ensemble''. 

We will show that the probability distribution of microstates generated by tapping agree with the prediction of the ``canonical ensemble'' only for intermediate tap intensities. For high and low tap intensities, the simulation explores only a portion of the phase space available.

Despite the shortcomings of the ensemble theory, this solvable model predicts a non-trivial maximum in fluctuations that can be observed in the simulations but only for the frictional case.

\section{Bowles-Ashwin model}

\begin{figure}
\begin{center}
\includegraphics[width=0.3\columnwidth]{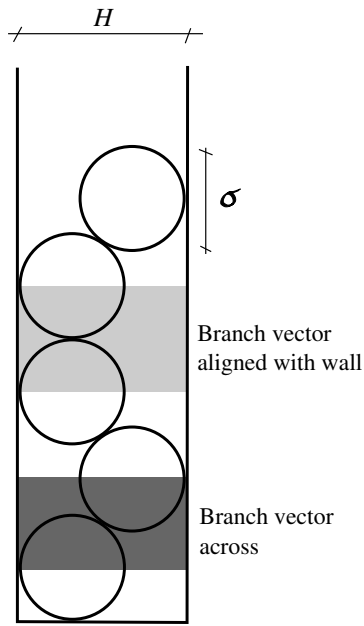}
\end{center}
\caption{Quasi-onedimensional model. The light shaded region corresponds to a branch vector aligned with the wall. The dark shaded area corresponds to a branch vector across the walls. Each configuration contributes with a distinct local volume to the total volume.} 
\label{model}
\end{figure}

\begin{figure}
\begin{center}
\includegraphics[width=0.32\columnwidth]{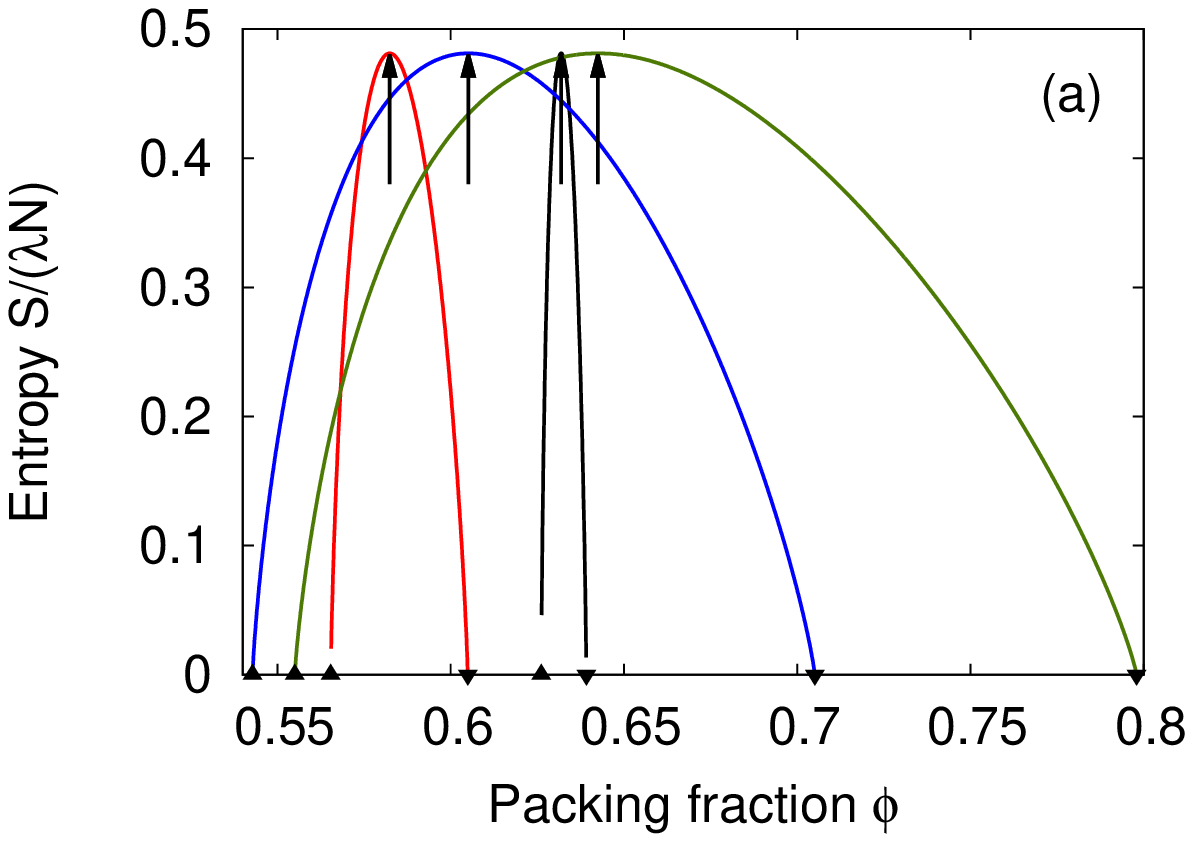}
\includegraphics[width=0.32\columnwidth]{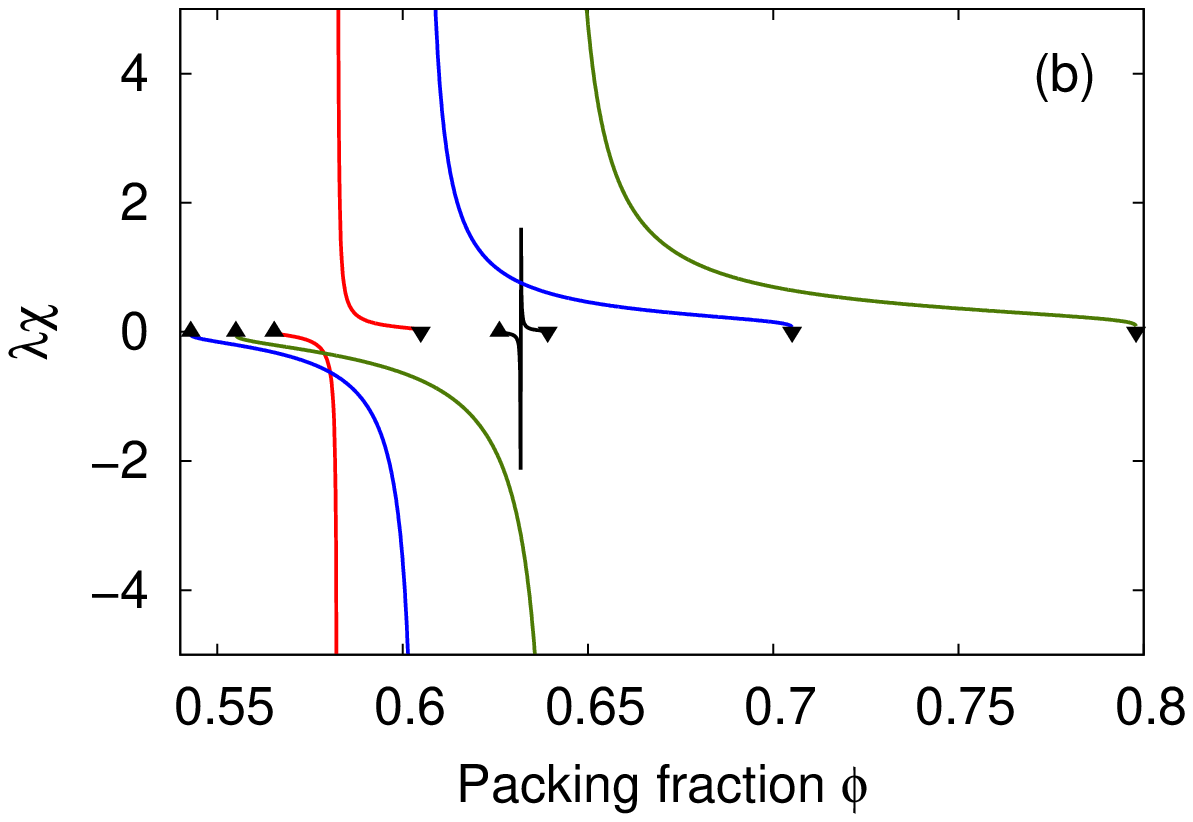}
\includegraphics[width=0.32\columnwidth]{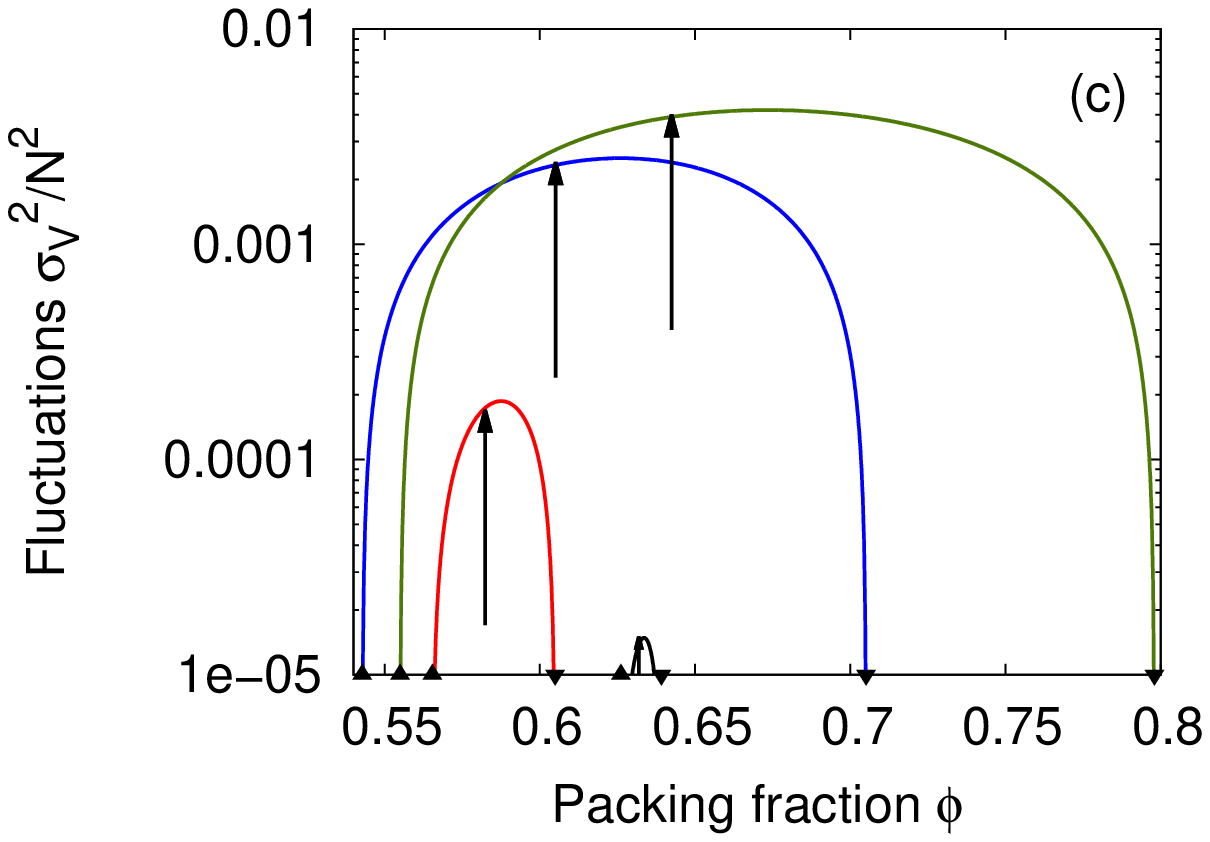}
\end{center}
\caption{(a) The entropy $S$ as a function of packing fraction $\phi$ for the Bowles-Ashwin model for $H=1.28$ (black), $1.48$ (red), $1.78$ (blue) and $1.846$ (green).  (b) The compactivity $\chi$. (c) The volume fluctuations $\sigma_V^2$. The black solid symbols indicate the position of the minimum (up triangles) and maximum (down triangles) $\phi$ allowed for each $H$. The corresponding packing fractions $\phi_{S_{max}}$ where entropy is maximum are indicated by arrows.} \label{entropy}
\end{figure}

We consider a model granular system introduced by Bowles and Ashwin which has an analytic density of states \cite{ashwin2009,bowles}. A set of frictionless disk of diameter $\sigma$ are arranged in a narrow channel of width $1<H/\sigma<(1+\sqrt{3/4})\approx1.866)$. A configuration mechanically stable against external pressure is obtained if the following local constraints are met. First, every disk needs three contacts, one with one of the walls of the channel and two with the only two possible neighbors (the possible  branch vectors for any disk are either aligned against a wall or across the channel). Second, the two branch vectors for a given disk cannot be both aligned against a wall (see Fig. \ref{model}). This allows to count all possible branch vector configurations (each has a well defined volume) using a simple binomial expression \cite{ashwin2009,bowles,ashwin2013}. As a result, the 
entropy $S(N,V)$ defined as the logarithm of the number of microstates associated with a given volume $V$ and number of disks $N$ is given by \cite{ashwin2013}.

\begin{equation}
\frac{S(N,V)}{\lambda N}=(1-\theta)\ln(1-\theta)-(1-2\theta)\ln(1-2\theta)-\theta\ln\theta,
\end{equation}  
where $\theta=M/N$, being $M<N/2$ the number of branch vectors aligned with the walls, and $\lambda$ plays teh role of the Boltzmann constant. The volume $V$ occupied by the disks for a given $\theta$ is $V=NH[\sqrt{(2-H)H}(1-\theta)+\theta]$. Notice that $V$ can only take discrete values since $\theta$ is a rational with $0<\theta<0.5$.

The compactivity $\chi$ is defined as the intensive variable conjugate to the volume, i.e., $\chi^{-1}=\partial S/\partial V$, hence

\begin{equation}
\chi=\frac{\partial S}{\partial \theta} \frac{\partial \theta}{\partial V} = \frac{H[1-\sqrt{(2-H)H}]}{\lambda [2\ln(1-2\theta)-\ln(1-\theta)-\ln(\theta)]}.
\end{equation}  

The volume fluctuations characterized by the variance $\sigma_V^2$ of the volume can be obtained from the ``specific heat'' as $\sigma_V^2=\lambda\chi^2\partial V/\partial \chi$. Therefore,

\begin{equation}
\sigma_v^2=\lambda\chi^2\frac{\partial V}{\partial \theta} \frac{\partial \theta}{\partial \chi} = \frac{N H^2\left[1-\sqrt{(2-H)H}\right]^2}{\left[\frac{4}{1-2\theta}-\frac{1}{1-\theta}+\frac{1}{\theta} \right]}.
\end{equation}  
 
In Fig. \ref{entropy} we show the entropy $S$, the compactivity $\chi$ and the volume fluctuations $\sigma_V^2$ as a function of the packing fraction $\phi=4\pi N/V$ for various values of $H$. Notice that the range of allowed $\phi$ varies with $H$. The entropy presents a maximum at $\phi_{S_{max}}$ ---in agreement with other models for static packings \cite{edwards,brey} and as estimated via simulations \cite{mcnamara,ciamarra}--- which corresponds to $\theta=1/2-\sqrt{5}/10$. States for packing fractions below $\phi_{S_{max}}$ correspond to negative compactivities associated to the inversion population of these volume bounded systems. Some authors suggest these negative $\chi$ macrostates may be inaccessible, however, this does not need to be the case; some preparation protocols may indeed lead to very low packing fractions \cite{ciamarra,pugnaloni2008}. Interestingly, fluctuations present a maximum at packing fractions above $\phi_{S_{max}}$. This maximum fluctuation should be then observed even if only 
positive $\chi$ states are assessed.

Notice, that the Bowles--Ashwin model is similar to the very first model proposed by Edwards \cite{edwards}. However, in Edwards simplistic model, particles are assigned one of two possible local volumes without restriction associated with the volumes already assigned to the neighbors as it is done here.

\section{Simulation}

\subsection{Model system}

We carry out discrete element simulations of disks of diameter $\sigma$ subjected to the gravity force $g$ and confined in a narrow container of width $H$. The container has a flat base and is infinitely high. The particle--particle and particle--wall interactions correspond to a normal restitution coefficient $\epsilon = 0.058$ and a static and dynamic friction coefficient $\mu_s = \mu_d = 0.5$ for the frictional simulations and $\mu_s = \mu_d = 10^{-5}$ for the ``frictionless'' simulations. In order to compare results from systems of different widths we vary the number $N$ of particles to ensure that a similar height of the granular column is obtained in all cases ($17\leq N \leq 24$). Simultaneously, we tune the material density of the disks in such way that all systems have the same total mass.

We used the Box2D library to solve the Newton-Euler equations of motion \cite{box2d}. Box2D uses
a constraint solver to handle hard bodies. At each time step of the dynamics a series of 25 iterations are used to resolve penetrations between bodies through a
Lagrange multiplier scheme \cite{catto}. After resolving penetrations, the inelastic collision at
each contact is solved and new linear and angular velocities are assigned to the particles. The equations of motion are integrated through a symplectic Euler algorithm. The time step $\delta t$ used is $0.0031$ $\sqrt{\sigma/g}$. Solid friction is also handled by means of a Lagrange multiplier
scheme that implements the Coulomb criterion. Previous works using this library have shown that simulations are consistent with other more complex interaction models for granular particles \cite{carlevaro2011,sanchez}.

\subsection{Tapping}

Tapping is simulated by setting the initial
velocity $v_0$ of the container (originally at rest) to a given positive value in the vertical direction. In doing so, the container, and the particles inside, moves upward and fall back on top of a zero restitution base. While the box dissipate all its kinetic energy on contacting
the base, particles inside the box bounce against the box walls and floor until they fully settle. After all particles come to rest a new tap is applied. The intensity of the taps is characterized by the initial velocity imposed to the confining box at each tap (i.e. $\Gamma = v_0$). 

The tapping protocol consists of a series of 46800 taps. Every 130 taps we change the value of $\Gamma$ by a small amount $\Delta\Gamma$. We initially decrease $\Gamma$ from about $20.0 (\sigma g)^{1/2}$ down
to a very low value and then increase it back to its initial high value. We repeat this protocol on the sample three times. At each value of $\Gamma$ the last 30 taps are used to average the packing fraction.

\section{Results}

\subsection{Packing fraction}

\begin{figure*}
\begin{center}
\includegraphics[width=0.4\columnwidth]{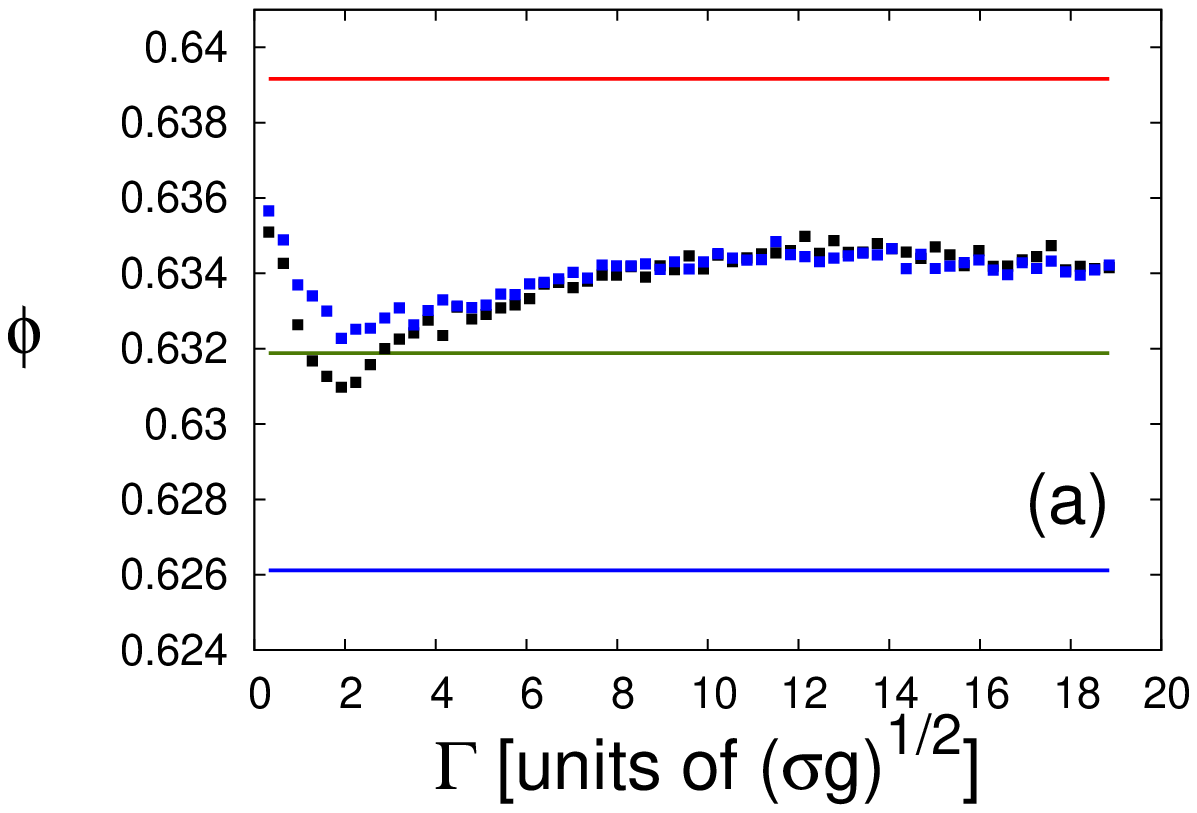}
\includegraphics[width=0.4\columnwidth]{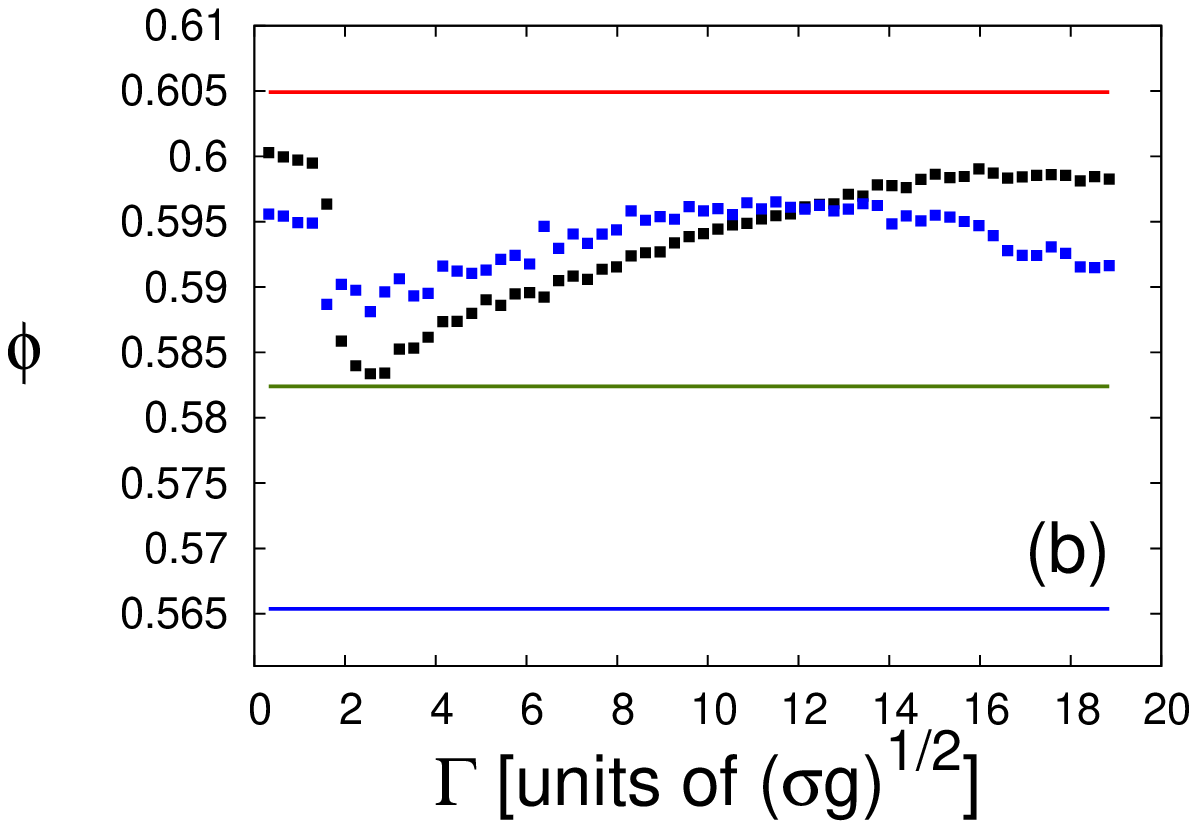}
\includegraphics[width=0.4\columnwidth]{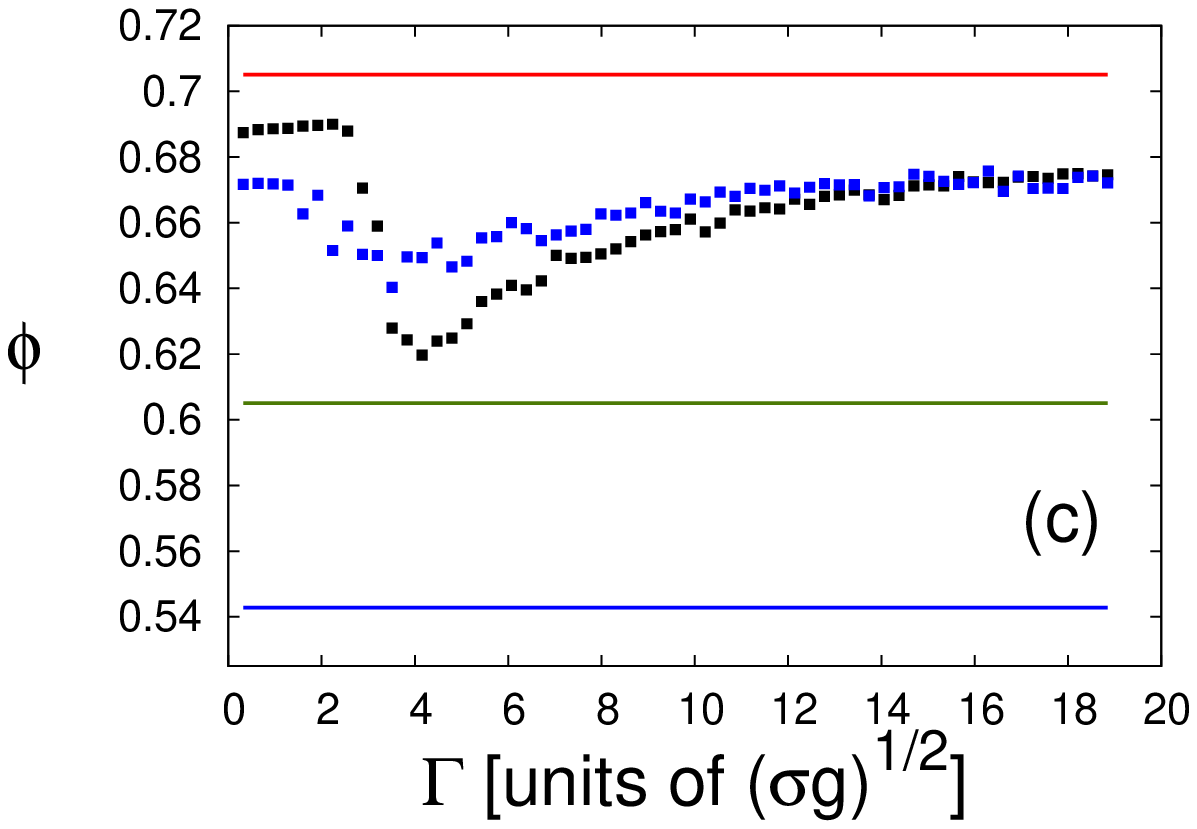}
\includegraphics[width=0.4\columnwidth]{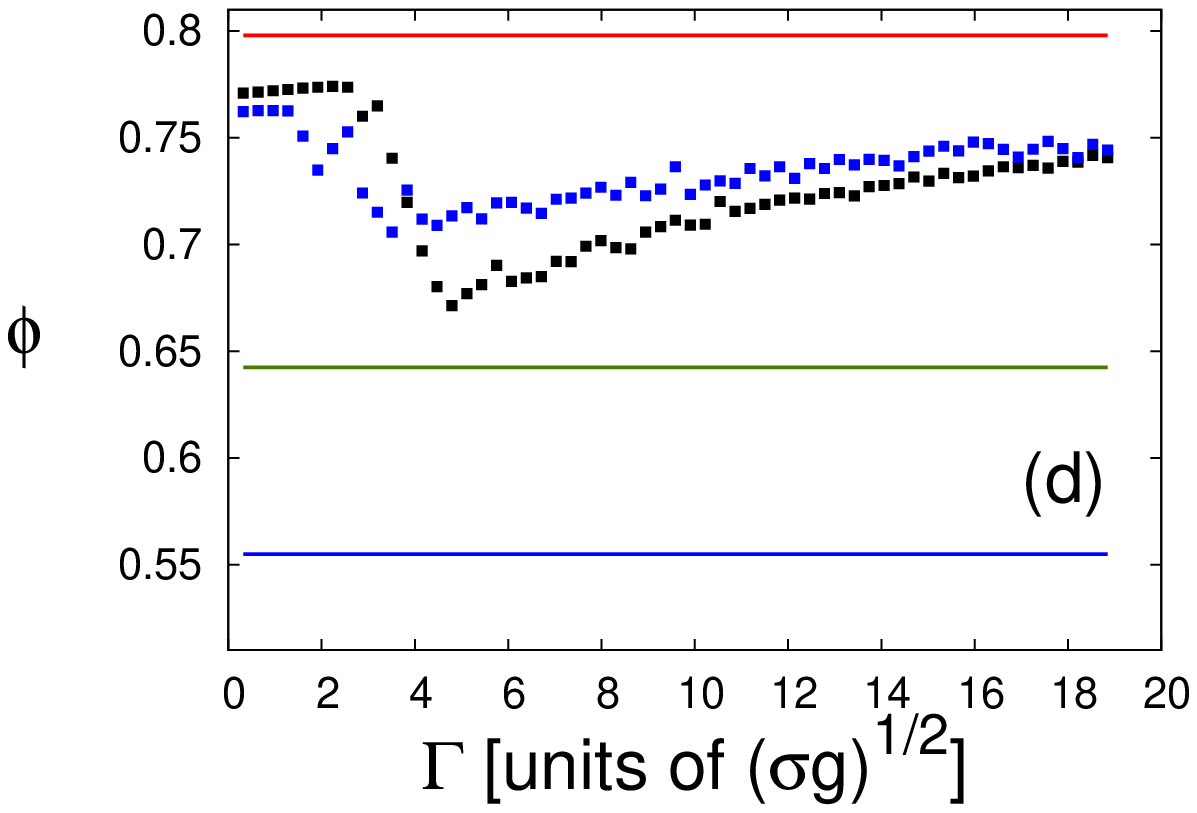}
\end{center}
\caption{The mean packing fraction $\phi$ as a function of tap intensity $\Gamma$ for frictional (black symbols) and frictionless (blue symbols) disks. (a) $H=1.28$, (b) $H=1.48$, (c) $H=1.78$, (d) $H=1.846$. The horizontal lines correspond to the maximum ($\phi_{max}$, red) and minimum ($\phi_{min}$, blue) packing fraction predicted by the Bowles--Ashwin model, and for the packing fraction ($\phi_{S_{max}}$, green) at which entropy is maximum.
} \label{phi}
\end{figure*}

In Fig. \ref{phi}, we show the mean packing fraction $\phi$ as a function of the tap intensity $\Gamma$ for various system widths $H$ for frictional and frictionless disks. The annealing protocol in which we start from high tap intensities yields a well defined mean packing fraction for any given $\Gamma$. Hence, we have averaged results from all three repetitions of the annealing.

Figure \ref{phi} shows results consistent with previous studies where a minimum packing fraction has been reported for wider 2D systems and 3D systems \cite{pugnaloni2010,pugnaloni2011,pugnaloni2008,carlevaro2011,gago}. The horizontal lines in Fig. \ref{phi} indicate the maximum and minimum values of $\phi$ allowed in the Bowles--Ashwin model. As we can see, tapping makes $\phi$ vary in a range much narrower than the possible theoretical values, both for frictional and frictionless disks. Frictional disks achieve higher and also lower packing fractions than frictionless disks. Interestingly, for $H=1.28$, frictional disks reach mean packing fractions below the one predicted to correspond to the maximum entropy $\phi_{S_{max}}$ indicating that negative $\chi$ may have been achieved (although we do not know the exact density of states for the frictional case). In all other cases, mean packing fractions remain above $\phi_{S_{max}}$.

\begin{figure}
\begin{center}
\includegraphics[width=0.5\columnwidth]{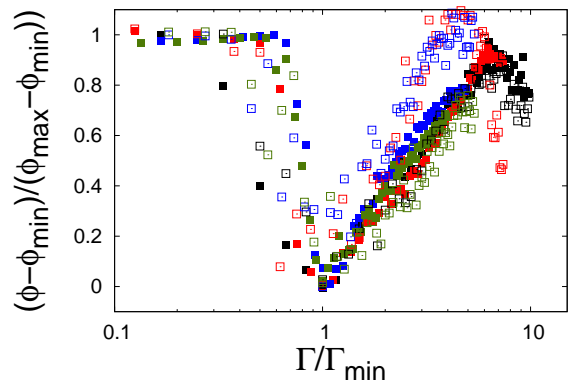}
\end{center}
\caption{Scaled $\phi'$ as a function of scaled $\Gamma'$ for frictional (solid symbols) and frictionless (open symbols) for $H=1.28$ (black), $1.48$ (red), $1.78$ (blue) and $1.846$ (green).} \label{scaled-phi}
\end{figure}

To compare results from different $H$, we plot in Fig. \ref{scaled-phi} the same data as in Fig. \ref{phi} scaled so that the minimum and maximum packing fraction in each curve corresponds to 0 and 1, respectively [i.e., $\phi'=(\phi-\phi_{min})/\phi_{max})$], and the position of the minimum $\phi$ also coincides (i.e., $\Gamma'=\Gamma/\Gamma_{min}$).  

The scaled $\phi'$--$\Gamma'$ curve shows a good collapse in the full range of tap intensities considered. This is an indication that, despite the large discrepancies in the range of densities achieve for different $H$, tapping induces a similar exploration of configurations in all these systems.

\subsection{Volume histograms}

\begin{figure}
\begin{center}
\includegraphics[width=0.32\columnwidth]{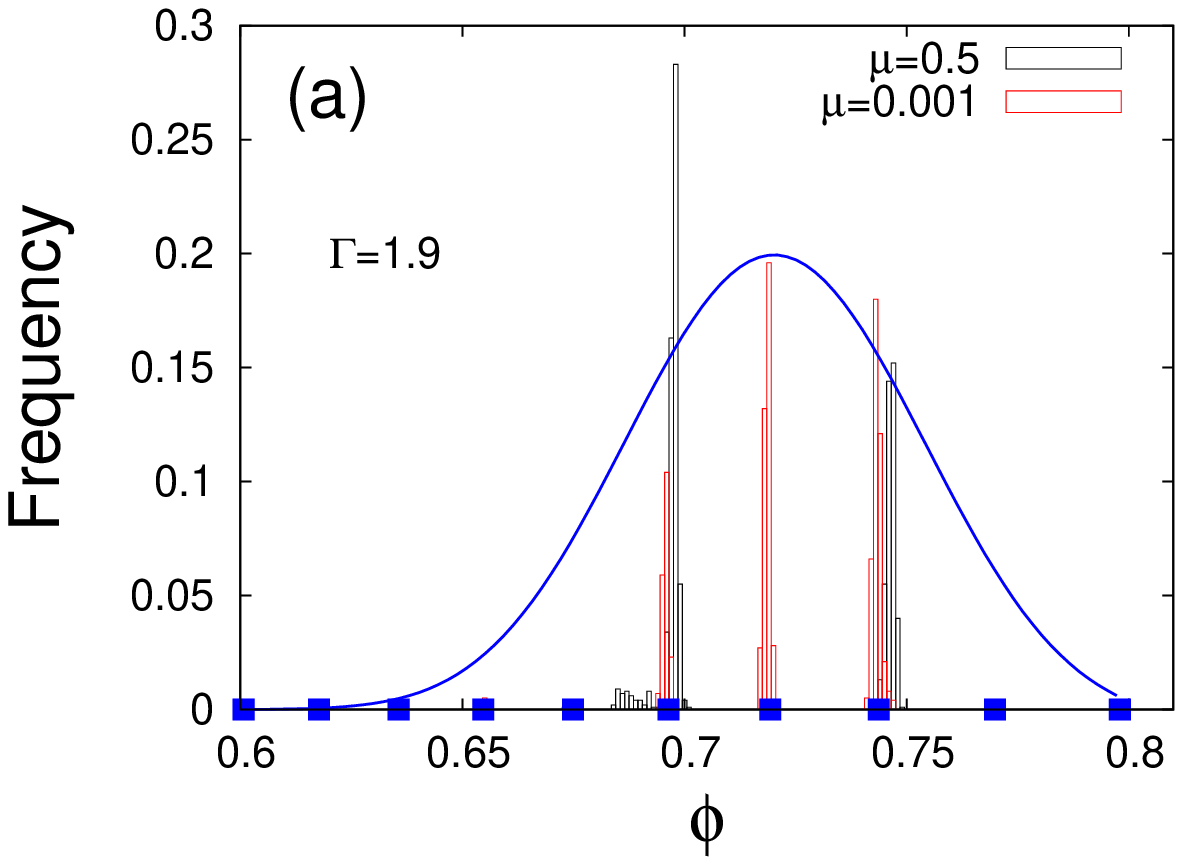}
\includegraphics[width=0.32\columnwidth]{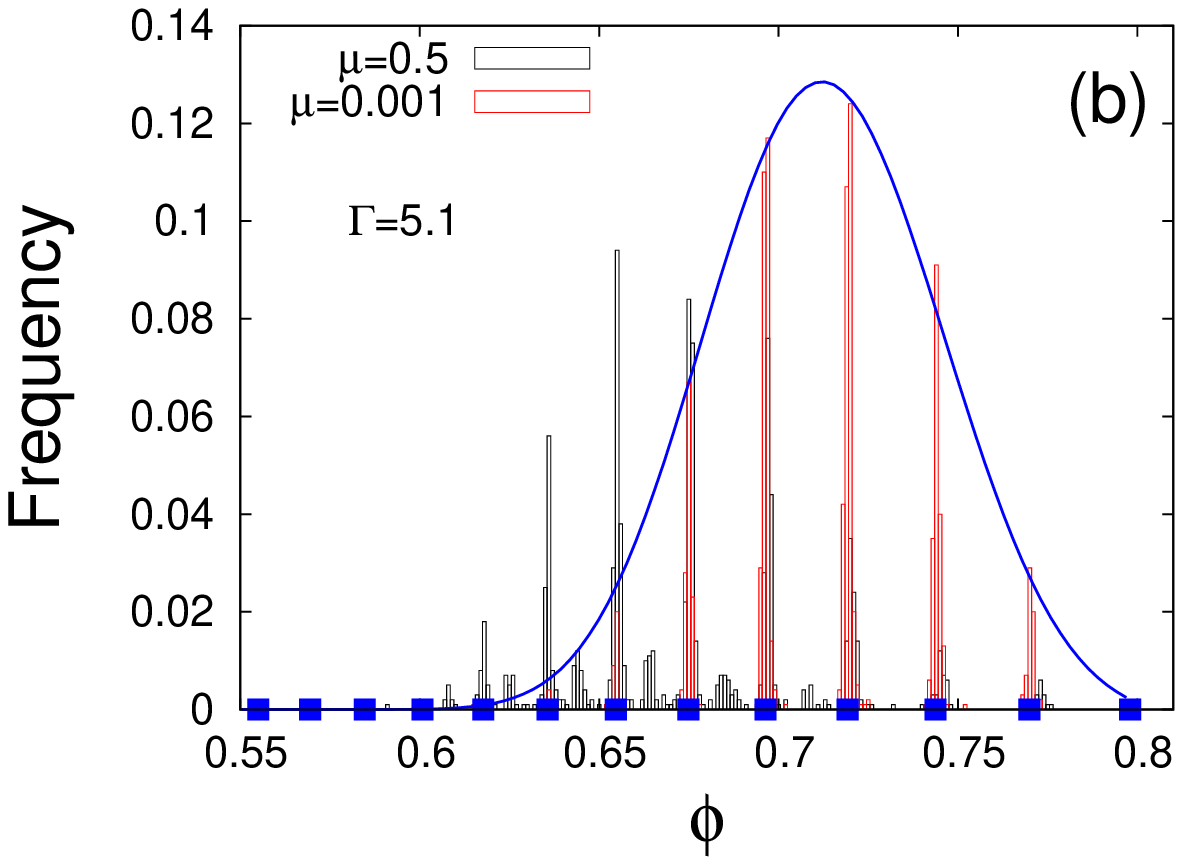}
\includegraphics[width=0.32\columnwidth]{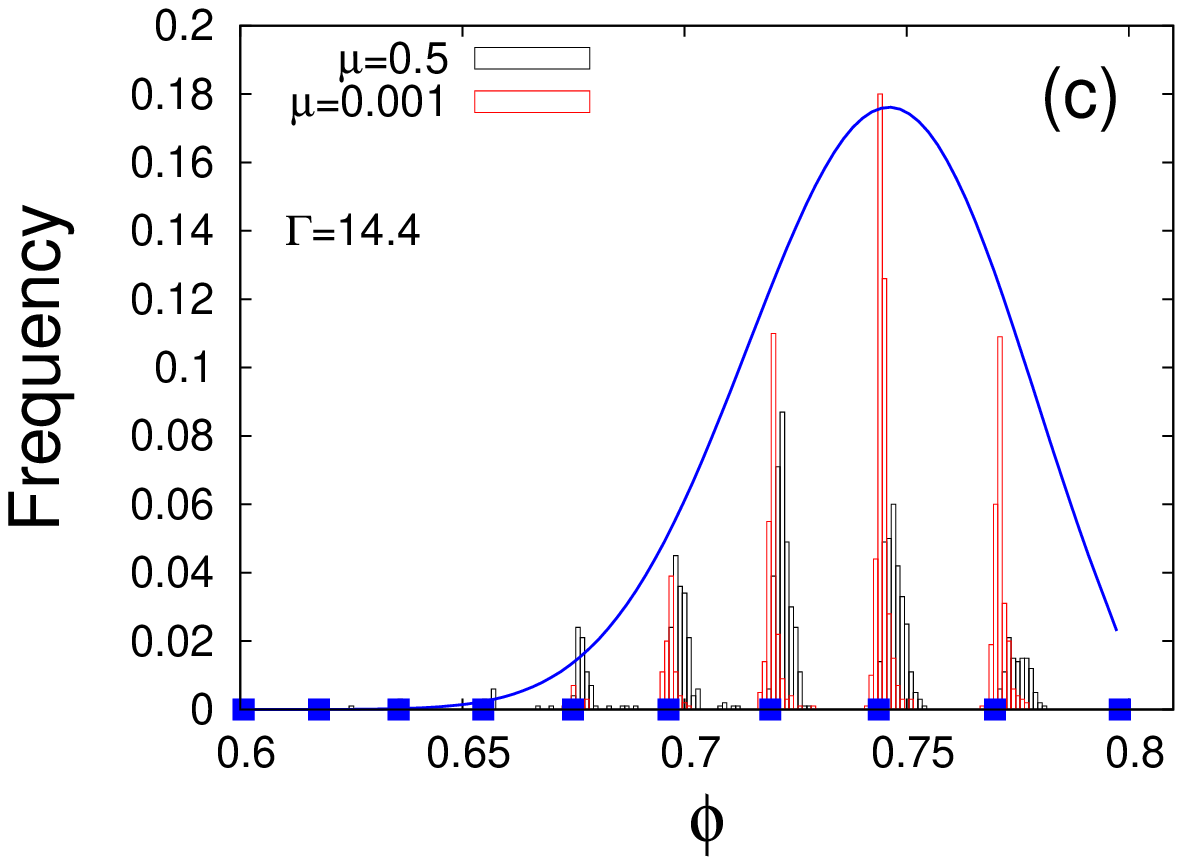}
\end{center}
\caption{Volume histograms for $H=1.846$ for frictional (black) and frictionless (red) disks. (a) $\Gamma=1.9$, (b) $\Gamma=5.1$, (c) $\Gamma=14.4$. The blue lines correspond to the analytical histogram $\propto \exp(S/\lambda)\exp[-V/(\lambda\chi)]$ where we have chosen $\lambda\chi$ to best fit the data for frictionless disks ($\Gamma=1.9 \rightarrow \lambda\chi=0.52$, $\Gamma=5.1 \rightarrow \lambda\chi=0.58$, $\Gamma=14.4 \rightarrow \lambda\chi=0.38$.). The blue squares indicate the discrete packing fractions (or volumes) allowed by the Bowles--Ashwin model for this 24-particle system.} \label{hist}
\end{figure}

The Bowles-Ashwin model considers frictionless disks and hence the mechanically stable configurations are reduced to a very small set. In Fig. \ref{hist}, we can see the distribution of volumes obtained in the simulations for $H=1.846$ at three different tap intensities. For these histograms, $1000$ taps in the steady state for the selected $\Gamma$ have been collected. For frictionless disks, the volume takes only the discrete values predicted by the model. However, frictional disks are able to arrange themselves in configuration having intermediate volumes since a disk may be stable with only two contacts thanks to friction even without contacting a wall. Despite this, frictional disks still take volumes predicted for frictionless disks in a much prominent fashion.

Since we know the density of states [$\exp(S/\lambda)$], the volume histograms in the ``canonical ensemble'' can be calculated as $\exp(S/\lambda)\exp[-V/(\lambda\chi)]$ \cite{edwards}. The blue lines in Fig. \ref{hist} represent this histogram with $\chi$ chosen as a fitting parameter. As we can see, low tap intensities yield histograms which are far from being described by the analogue of the system being coupled to a ``volume bath'' of given $\chi$. For higher tap intensities, the assumption that tapping takes the system to a macroscopic state compatible with a bath at fixed $\chi$ is fair. However, notice that high tap intensities lead to low $\chi$ and the system is unable to sample the smallest possible volume (highest packing fraction) in our simulations which has a non-zero probability according to the theory [see Fig. \ref{hist}(c)].

\subsection{Volume fluctuations}

\begin{figure*}
\begin{center}
\includegraphics[width=0.4\columnwidth]{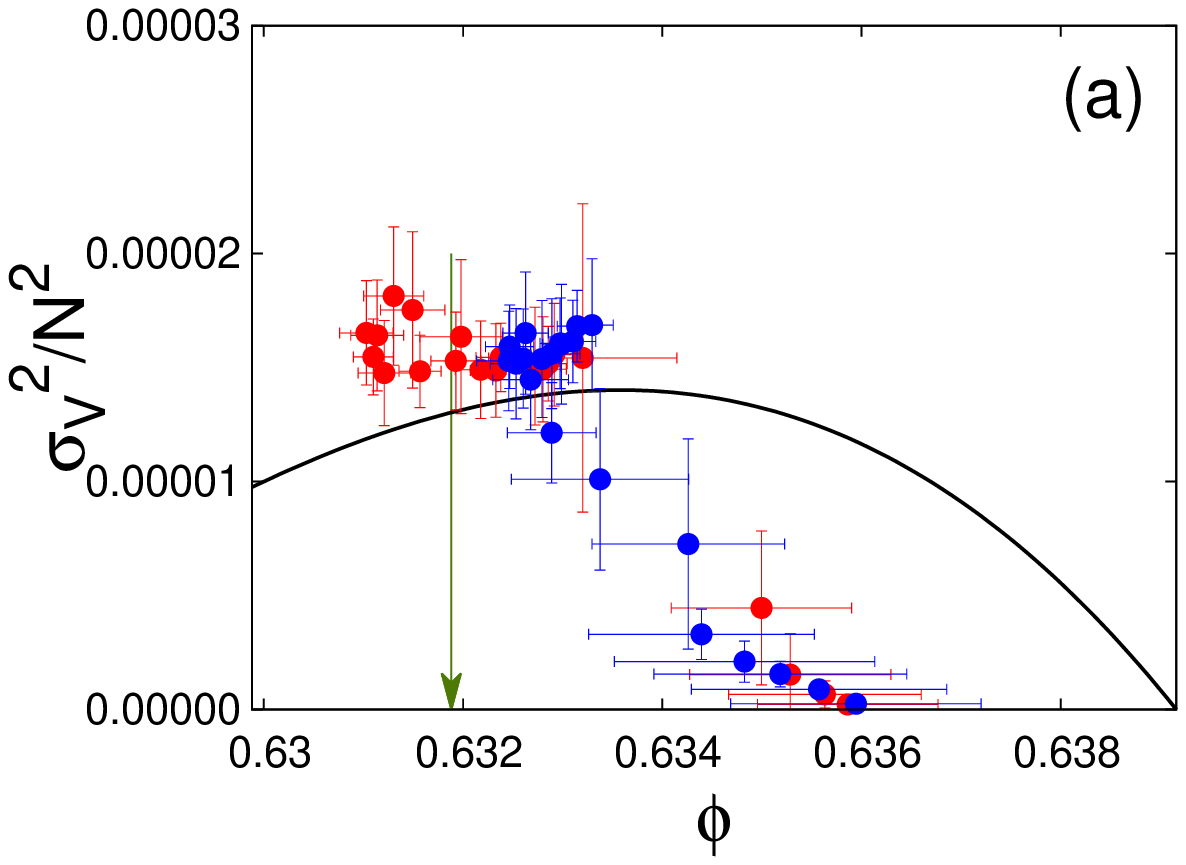}
\includegraphics[width=0.4\columnwidth]{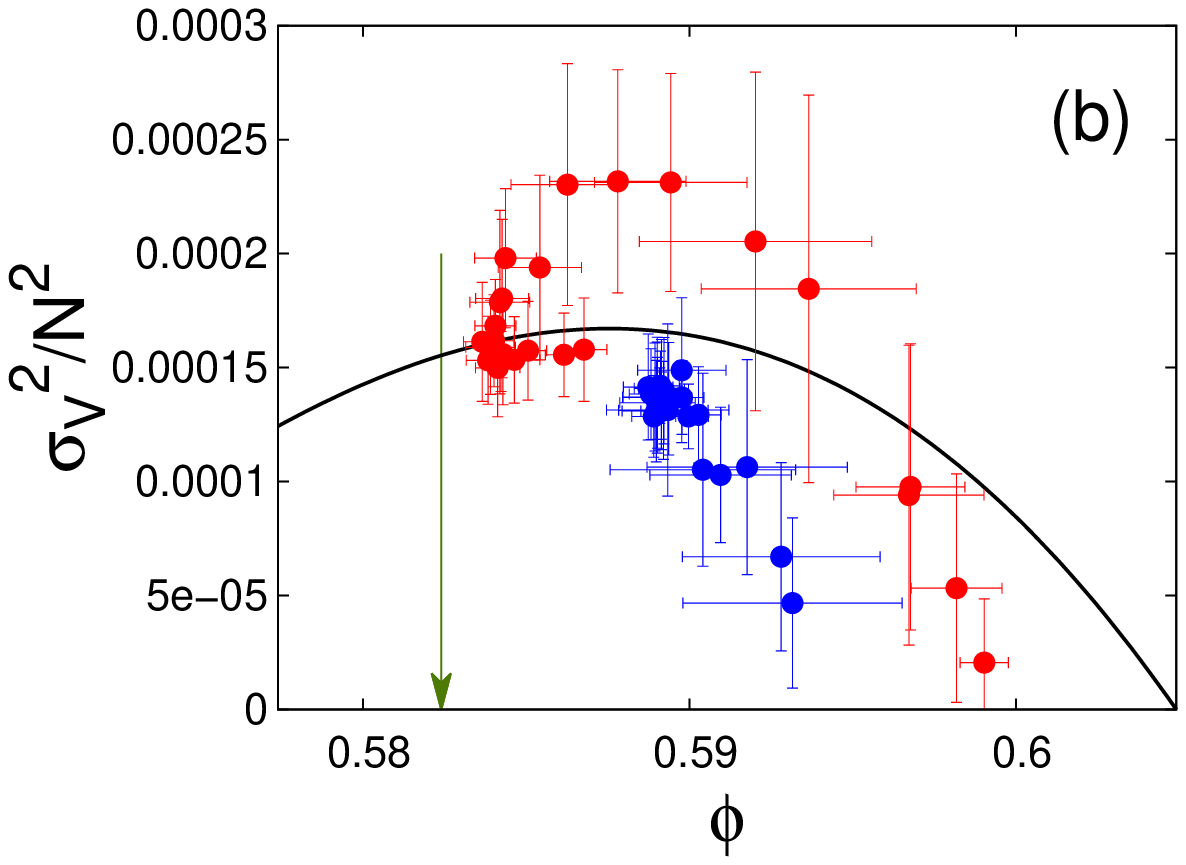}
\includegraphics[width=0.4\columnwidth]{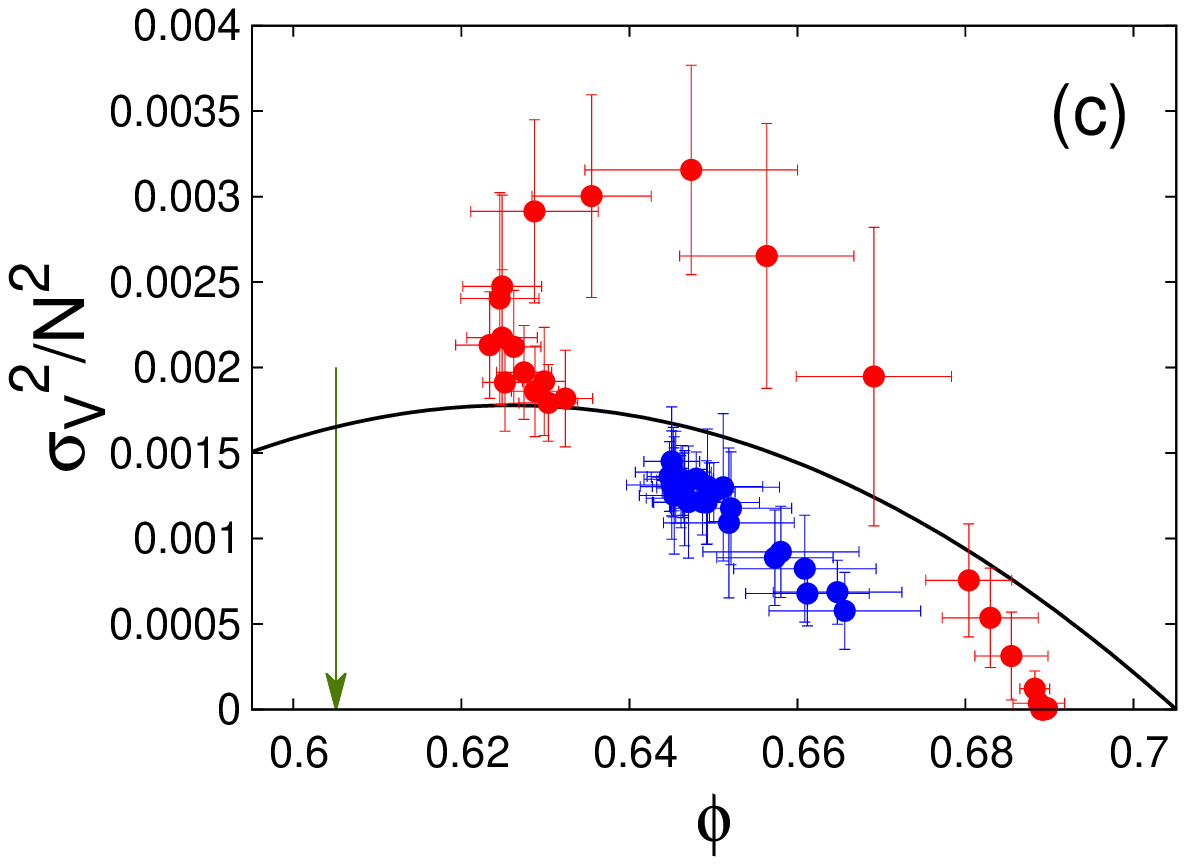}
\includegraphics[width=0.4\columnwidth]{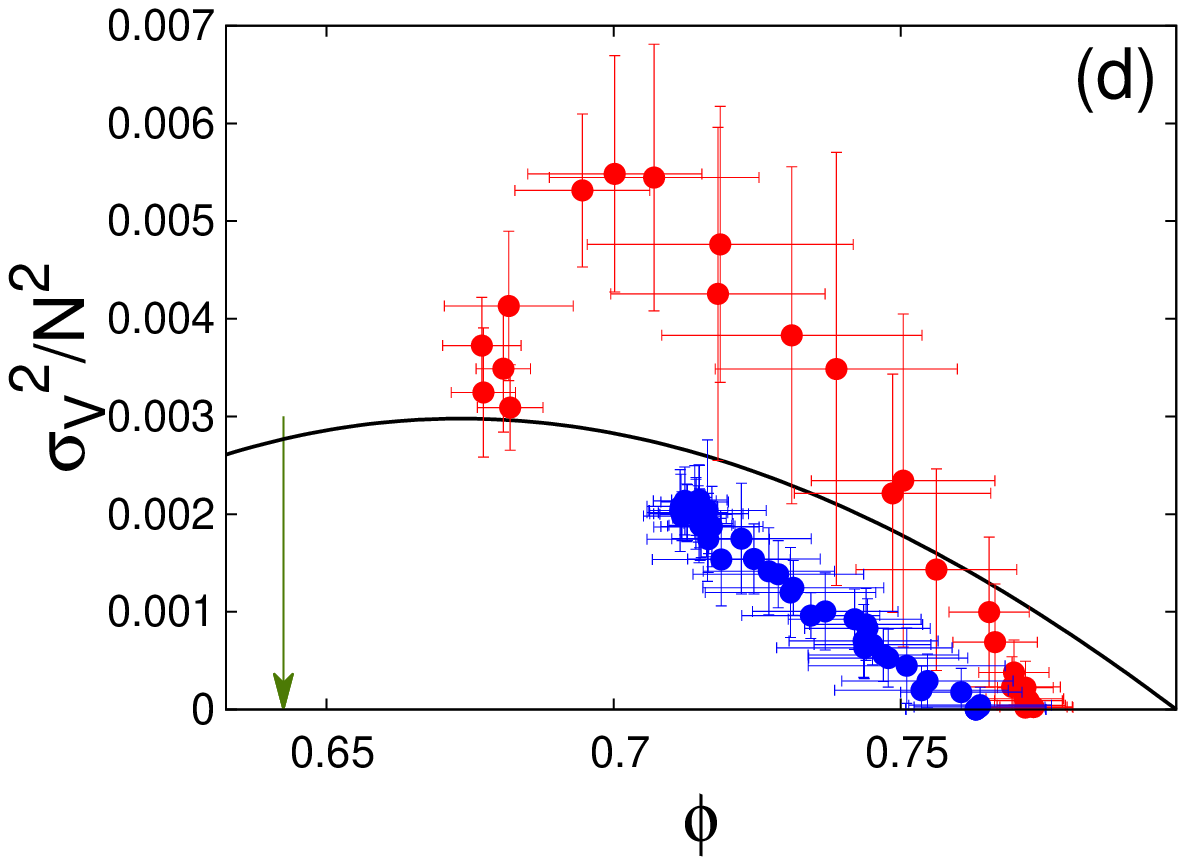}
\end{center}
\caption{Fluctuations of volume for frictional (red) and frictionless (blue) disks for $H=1.28$ (a),  $1.48$ (b), $1.78$ (c), and  $1.846$ (d). Error bars were obtained as the standard deviation over 20 independent instances of the simulation. Lines correspond to the analytic prediction for the Bowles-Ashwin model. The arrows indicate the position $\phi_{S_{max}}$ of the maximum entropy state.} \label{fluct}
\end{figure*}

Figure \ref{fluct} shows the volume fluctuations $\sigma_V^2$ as a function of packing fraction for various widths of the container. We have increased the statistics in Fig. \ref{fluct} by running 20 independent instances to estimate error bars in the $\Gamma$ region of interest.  As we discussed in section 2, the Bowles--Ashwin model predicts a maximum in the fluctuations for a packing fraction above $\phi_{S_{max}}$. It is clear form Fig. \ref{fluct} that fluctuation present a maximum for the frictional disks. In contrast, for $H>1.28$, frictionless disks under tapping sample higher values of $\phi$ where a monotonic decrease in $\sigma_V^2$ is predicted. For $H=1.28$, frictionless disks do sample states at $\phi$ low enough for the maximum in fluctuations to be observed. Although subtle, it can be seen that fluctuations grow slightly up to $\phi \approx 0.633$ and decay quickly beyond that value.

The maximum in the fluctuations has been observed previously in quasi-2D experiments of tapped granular columns \cite{pugnaloni2010,pugnaloni2011}. However, contrasting results have been found in 2D and 3D packings \cite{ciamarra2006,puckett,briscoe} where a monotonic decay with increasing packing fraction has been observed, whereas in 3D fluidized beds a minimum in the fluctuations was reported \cite{schroter}. While dimensionality and jamming state may be the reasons behind such strong qualitative discrepancies between experiments, it is clear that in the present system, both, the analytic ensemble theory and the DEM simulation of tapping agree in that a maximum in fluctuations exists. 

Figure \ref{fluct} makes apparent an additional feature of fluctuations in the case of frictional disks. Different fluctuations are observed even if the mean volume of the macrostate is the same [see parts (b) and (c)]. This has been observed before in similar systems \cite{pugnaloni2010,pugnaloni2011}. The implication is that the macrostates are not fully described by the volume and further macroscopic variables are needed. In Refs. \cite{pugnaloni2010,pugnaloni2011} it was concluded that adding the force moment tensor as an extra extensive variable allowed the states to be fully described having same fluctuations if the same mean volume and mean force moment tensor where observed for two states obtained through different tap intensities and tap duration.

\section{Conclusions}

We have tested the prediction of the Edwards ensemble for the Bowles--Ashwin model by running DEM simulations of tapped columns of frictionless and frictional disks. These simulations of a realistic protocol of generating steady states for a model with analytic solution is particular suited to validate the ensemble approach to describe the statistics of static packings. One interesting finding is that fluctuations are predicted by the ensemble theory to display a maximum as a function of $\phi$ and this has been indeed observed in the simulations.

Beyond the overall similarity between the theoretical and simulation results, there are clear differences that suggest the ensemble theory is unable to capture some of the response of the realistic system. In particular, the range of volumes explored through tapping is much narrower than the full range of allowed volumes. This feature akin to ergodicity breaking \cite{paillussion}. Tapping may condition the system to explore only a region of the phase space. However, we find that independent instances of the tapping protocol do not lead to sampling a different portion of the phase space as expected in ergodicity breaking systems such as glasses. This would imply that the flat measure in the volume ensemble proposed initially by Edwards is inadequate at least for this system (but possibly for most granular systems). A revision may be necessary to consider 
that each type of perturbation would induce the system to sample the phase space in a different way.

\ack
L.A.P. acknowledges The Aspen Center for Physics where parts of this work were done, and S. S. Ashwin for valuable discussions.

\end{document}